\documentclass[12pt,preprint]{aastex}

\newcommand{\myemail}{courteau@astro.ubc.ca}
\newcommand{\etal}{et~al.~}

\newcommand{\hunit}{km~sec$^{\hbox{\scriptsize 
            -1}}$~Mpc$^{\hbox{\scriptsize -1}}$}
\newcommand{\hub}{H$_{\hbox{\scriptsize 0}}$}
\newcommand{\kms}{\ifmmode\,{\rm km}\,{\rm s}^{-1}\else km$\,$s$^{-1}$\fi}
\newcommand{\magarc}{\ifmmode {{{{\rm mag}~{\rm arcsec}}^{-2}}}
             \else {{{mag}$~${arcsec}$^{-2}$}}
             \fi}

\def \spose#1{\hbox to 0pt{#1\hss}}
\def \lta{\mathrel{\spose{\lower 3pt\hbox{$\sim$}}
     \raise 2.0pt\hbox{$<$}}}
\def \gta{\mathrel{\spose{\lower 3pt\hbox{$\sim$}}
     \raise 2.0pt\hbox{$>$}}}

\def\Equ#1{Eq.~(\ref{eq:#1})}
\def\se#1{\S\ref{sec:#1}}
\def\Fig#1{Fig.~\ref{fig:#1}}

\def\eg{{e.g.~}}

\def\be{\begin{equation}}
\def\ee{\end{equation}}
\def\prop{\propto}
\def\ifm#1{\relax\ifmmode#1\else$\mathsurround=0pt #1$\fi}
\def\kms{\ifmmode\,{\rm km}\,{\rm s}^{-1}\else km$\,$s$^{-1}$\fi}

\def\lsol{L_{\odot}}

\def\ltsima{$\; \buildrel < \over \sim \;$}
\def\lsim{\lower.5ex\hbox{\ltsima}}
\def\gtsima{$\; \buildrel > \over \sim \;$}
\def\gsim{\lower.5ex\hbox{\gtsima}}

\def\omm{\Omega_{\rm m}}
\def\omb{\Omega_{\rm b}}

\def\Vv{V_{\rm vir}}
\def\Vobs{V_{\rm obs}}

\def\Mv{M_{\rm vir}}
\def\Rv{R_{\rm vir}}

\def\fbar{f_{\rm bar}}

\def\Ld{L}
\def\Rd{R_{\rm d}}
\def\Md{M_{\rm d}}
\def\Mh{M_{\rm h}}

\def\Re{R_{\rm eff}}
\def\SBe{\Sigma_{\rm eff}}

\def\gf{\epsilon_{\rm gf}}

\def\mus {\mu_0}

\def\Upstilde{\widetilde{\Upsilon}}
\def\Upsilond{\Upsilon_{\rm d}}

\def\pmb#1{\setbox0=\hbox{#1}%
\kern-.025em\copy0\kern-\wd0
\kern.05em\copy0\kern-\wd0
\kern-.025em\raise.0433em\box0}

\def \ion#1#2{#1{\footnotesize{#2}}\relax}
\def \ha       {H$\alpha$\ }
\def \hi       {\ion{H}{I} }

\def \farcm{\hbox{$.\mkern-4mu^\prime$}}

\def \Rx        {R_{\rm d}}
\def \V22{V_{2.2}}
\def \r22{R_{disk}}

\def \logV      {\log{V}}

\def \logL      {\log{L}}
\def \dlogV     {\partial\log{V}}
\def \dlogR     {\partial\log{R}}
\def \dlogL     {\partial\log{L}}
\def \partialvr {\dlogV \thinspace / \thinspace\dlogR}
\def \partialVRL {\dlogV{(L)} \thinspace / \thinspace\dlogR{(L)}}
\def \partialRLV {\dlogR{(V)} \thinspace / \thinspace\dlogL{(V)}}
\def \partialVLR {\dlogV{(R)} \thinspace / \thinspace\dlogL{(R)}}
\def \littlemm{\ifmmode{\scriptscriptstyle m }
     \else{\hbox{$\scriptscriptstyle m $ }}\fi}
\def \topemm{\raise .9ex \hbox{\littlemm}}
\def \magpoint{\hbox to 2pt{}\rlap{\hskip -.5ex \topemm}.\hbox to 2pt{}}
\def \deg {$^\circ$}

\shorttitle{Scaling Relations of Local Spiral Galaxies}
\shortauthors{Courteau et~al.}

\slugcomment{Submitted to the Astrophysical Journal}
\received{2003 October XX}
\begin{document}

\title{Scaling Relations of Local Spiral Galaxies.I. Observational Foundations}

\author{St\'{e}phane Courteau and Lauren MacArthur}

\affil{Dept. of Physics \& Astronomy, University of British Columbia,
       6224 Agricultural Road, Vancouver, BC V6T 1Z1}
\email{\myemail, lauren@astro.ubc.ca}

\author{Avishai Dekel}
\affil{Racah Institute of Physics, The Hebrew University, Jerusalem 91904, 
       Israel}
\email{dekel@astro.huji.ac.il}

\author{Frank van den Bosch}
\affil{Department of Physics, Swiss Federal Institute of
       Technology, ETH H\"onggerberg, CH-8093, Zurich,
       Switzerland}
\email{vdbosch@phys.ethz.ch}

\author{Daniel H. McIntosh}
\affil{University of Massachusetts, Department of Astronomy, Amherst, 
      MA, 01003}
\email{dmac@hamerkop.astro.umass.edu}

\and 

\author{Daniel Dale}
\affil{Department of Physics \& Astronomy, University of Wyoming,
       Laramie, WY 82071}
\email{ddale@uwyo.edu}

\bigskip

\begin{abstract}
This paper presents an exploration of two fundamental scaling
relations of spiral galaxies, the luminosity-rotation speed
(or Tully-Fisher; TF) relation, and the size-luminosity (SL)
relation, and the dependences of their scatter, at red and 
infrared bands. We verify that the observed virial 
relations of disk galaxies are given by $\Vobs \prop L_I^{0.31}$
and $\Rx \prop L_I^{0.33}$ using distance-redshift surveys of 
high surface brightness (HSB) and low surface brightness (LSB) 
non-interacting galaxies.  These results imply that the galaxy 
surface brightness $\Sigma \propto \Rd \propto \Vobs \propto 
\Mv^{1/3}$.  The collected surveys provide accurate $I$-band 
luminosities, disk scale lengths, circular velocities, and, in 
some cases, color. Various issues regarding the scatter of 
scaling relations in blue to near-infrared bands are being 
re-examined with accurate $JHK$ Kron luminosities, effective 
radii, and colors from the 2MASS database.
We derive the first extensive $J$-band TF and SL relations.  
At a given infrared luminosity, the TF velocity residuals are
correlated with infrared color, which in turn is determined by the variations 
in galaxy formation ages and dark halo concentrations; these residuals 
are fully 
independent of surface brightness and other tested galaxy observables.  
We verify that TF relations (TFRs) for various morphological types have 
different 
zero points, but a common slope, such that early-type (redder) disks 
rotate faster than later-type (bluer) systems of the same luminosity. 
The morphological type dependence of the TFR is a direct consequence of 
the more fundamental scatter dependence on color, which itself is related 
to the star formation history of a galaxy.  The scatter of the SL relation 
is mostly dominated by surface brightness, but color also plays a small role.  
The observed systematic variations of disk size with color,
for a given luminosity, are weaker than expected, perhaps as a 
result of uncertainty in disk scale length measurements. 
The TF and SL residuals for HSB and LSB galaxies are weakly correlated 
with $\partialVRL =-0.07 \pm 0.05$, in agreement with the earlier 
claim by Courteau \& Rix (1999, ApJ, 513, 561), and
$\partialRLV = \phantom{-}0.12 \pm 0.03$.  The former result
suggests that spiral disks of all color, brightness and barredness 
may be, on average, dominated by dark matter even in their inner parts 
(the so-called ``sub-maximal disk'' solution at $R \gta 2.2$ disk
 scale lengths). 
The observed scaling relations of disk galaxies are likely the result 
of the simplest scenario of galaxy formation in the standard cosmological 
picture, but their color dependence remains a challenge to hierarchical 
structure formation models.  These relations and their derivatives are 
analysed in terms of, and serve as stringent constraints for, galaxy 
formation models in our companion paper. 

\end{abstract}


\section{Introduction}\label{sec:intro}

Understanding the origin and nature of galaxy scaling relations 
is a fundamental quest of any successful theory of galaxy formation.
The success of a particular theory will be judged by its ability to
predict the slope, scatter, and zero-point of any robust galaxy
scaling relation at any particular wavelength.  Some observed scaling
relations in spiral galaxies, based on their size, luminosity, and 
rotation speed, can be reproduced {\it individually} to fairly good accuracy
by invoking galaxy formation models that include virial equilibrium
after dissipational collapse of spherical cold dark matter (CDM)
halos and angular  momentum conservation (e.g. Mo, Mao, \& White 1998, 
hereafter  MMW98; van den Bosch 1998, 2000, hereafter collectively 
as vdB00; Navarro \& Steinmetz 2000, hereafter NS00; Firmani \& 
Avila-Reese 2000; hereafter FAR00). 

One of the most 
firmly established empirical scaling relation of disk galaxies is the 
Tully-Fisher relation (TFR; Tully \&  Fisher 1977); a tight correlation 
between the total luminosity and the rotation speed of a disk galaxy.  
However, to date, no single CDM-based model of galaxy formation 
can {\it simultaneously} reproduce the slope, zero-point, scatter, and color 
trends of the TFR, match the shape 
and normalization of the luminosity function, and explain the sizes, colors, 
and metallicity of disk galaxies (see, e.g,. vdB00; Bell~\etal 2003).  
In addition, simultaneously accounting for the mass and angular momentum 
distribution of spiral galaxies in a gas dynamical context remains a 
major challenge for hierarchical formation models (Navarro \& White
1994; Bullock \etal 2001b; van den Bosch \etal 2002b).
A complete theory of galaxy scaling relations awaits a fuller understanding 
of structure forming mechanisms and evolutionary processes (e.g. star 
formation, 
merging, feedback, and cooling prescriptions) in galaxies (see Somerville 
\& Primack 1999 for a comparison of ``recipes'' among competing theories.)  
Likewise, the fine-tuning of these galaxy formation and evolutionary models 
demands a careful examination of empirical scaling relations of galaxies.  

In order to set up a framework for the study of galaxy scaling relations, 
we examine the correlations of parameters related by the virial theorem, 
$V^2 \prop M/R$, for bright galaxies (i.e. where feedback effects are minimal).
We consider three fundamental observables for each disk galaxy: the total 
luminosity $L$, the stellar scale length $\Rd$ of the exponential disk, 
and the observed circular velocity $\Vobs$. 
The stellar mass, $M_d$, can be estimated from the luminosity by assuming a 
stellar mass-to-light ratio, $\Upsilond=M_d/L$.  The size-luminosity (SL) 
relation of galaxy disks is also expressed as $L \propto \Sigma \Rd^2$, 
where $\Sigma$ is the surface brightness.

Key to mapping fundamental dynamical trends in spiral galaxies, the 
measurement of TF and SL relations and detection of their correlated 
residuals require velocity amplitudes measured at a suitably chosen 
radius representative of the flat part of resolved rotation curves, 
red/infrared magnitudes to minimize extinction and 
population effects, accurate disk scale lengths, and, ideally, color 
terms with a broad baseline (e.g. $B-K$) to test for $\Upsilond$ 
variations in the stellar population and extinction effects.  
Our goal in this paper is thus to assemble such a data base.

A study of scaling relations in irregular and spiral galaxies by
Salpeter \& Hoffman (1996; hereafter SH96) yielded the correlations
$L_B \propto R^{2.68} \propto \Vobs^{3.73} \propto M_H^{1.35} \propto 
M_{dyn}^{1.16}$, where $M_H$, $M_{dyn}$, and $R$, are the \hi and 
dynamical masses, and a characteristic radius, respectively.  The 
blue luminosities, as used
in that study, are notoriously sensitive to dust extinction and stellar 
population effects and dynamical effects cannot be simply isolated.  
A new study of scaling relations in the (near-)infrared would 
provide more robust dynamical constraints to galaxy formation models. 
While the TFR has been examined at nearly all optical-IR wavelengths
(e.g. Strauss \& Willick 1995; Verheijen 2001, hereafter V01), 
comparatively few multi-wavelength analyses of the SL relation (SLR) of 
spiral galaxies have been done so far (SH96; MMW98; Shen \etal 2003).   
This is partly because the accurate disk scale lengths needed to calibrate 
the SLR, at any wavelength, have only recently become available 
for large databases (e.g. Courteau 1996; Dale \etal 1999; MacArthur 
\etal 2003, hereafter MCH03).  The recent availability of light profile 
decompositions for very large galaxy databases (e.g. Sloan Digital
Sky Survey and Two Micron All-Sky Survey; \se{data}) heralds a new 
era for the study of galaxy scaling relations (e.g. Shen \etal 2003).  
We will return to considerations about the SLR in \se{virial}.  The 
remainder of this section focuses on the TFR and the dependences 
of its scatter on galaxy observables.

The empirical TFR is expressed as
\be 
L \propto V^a_{\rm obs}
\label{eq:IRTF}
\ee
with the near-IR log-slope $a \simeq3.0 \pm 0.4$ (Willick \etal 1997; 
Giovanelli \etal 1997, hereafter G97; Courteau \etal 2000; V01). 
Reported values of the log-slope $a$ range from 2.8 in the blue to 4.0
in the infrared (Willick \etal 1997; Tully \& Pierce 2000; V01) for 
both high and low surface brightness galaxies
(Zwaan \etal 1995; V01). Log-slopes steeper than $a \sim 3.5$ in the
infrared typically result from small samples and excessive pruning on 
the basis of idealized morphology or kinematics, a narrow
range of inclinations, redshift cutoffs, etc.  (Bernstein \etal 1994;
V01; Kannappan, Fabricant, \& Franx 2002, hereafter KFF02). The slope,
scatter, and zero-point of blue TFRs are predominantly dominated by
stellar population and dust extinction effects (e.g. Aaronson \& Mould
1983; C97; G97; Willick \etal 1997; Tully \& Pierce
2000) and on the techniques used to recover the major observables and
fitting for fundamental relations (e.g. Strauss \& Willick 1995; C97;
V01; Bell \& de Jong 2001; KFF02).
Because we are mainly interested in masses, rather than luminosities, 
we do not concern ourselves with TFRs and other scaling relations
measured at blue wavelengths.  We show in Appendix A that the fundamental 
form of the TFR, based on dynamical principles only, is given by $a=3$, 
that is $L \propto \Vobs^3$.

The modern interpretation of the TFR is that of a correlation between
the total baryonic mass of a galaxy, inferred via its infrared
luminosity and a stellar mass-to-light ratio and total gas mass (\hi +
He), and its total mass inferred  from the  asymptotic circular
velocity of the  galaxy disk (McGaugh \etal 2000; Bell \& de Jong 2001;
V01). The ``baryonic'' TFR is expressed as 
\be
{\cal M}_{\rm{bary}} \propto V^{a_{\rm bary}}_{\rm obs}
\label{eq:barTF}
\ee

Since disk gas mass fractions typically increase with decreasing
luminosity (i.e., McGaugh \& de Blok 1997), one typically has that
$a_{\rm bary} < a$.

While the log-slope of the TFR can be reproduced fairly well by most 
CDM-based structure formation  models (e.g. MMW98; vdB00; NS00), 
the predicted scatter can be large compared  to the inferred
``cosmic'' scatter of $\lsim 0.25$ mag in red/infrared bands
(Willick \etal 1996; V01) and interpretations about its dependence
differ (see below).

Besides the basic understanding of the slopes of galaxy scaling relations, 
the dependence of their scatter has also been addressed by many, especially
for the TFR  (e.g. Aaronson \& Mould 1983; Giraud  1986; Rhee 1996; 
Willick \etal 1997; KFF02), and can be used to set stringent constraints on 
structure formation models (Courteau \& Rix 1999, hereafter CR99; 
Heavens \& Jimenez 1999; FAR00; NS00; V01; Buchalter, Jimenez, \& Kamionkowski
2001; Shen, Mo, \& Shu 2002).  

While various trends in the scatter of the blue TFR have been reported in the 
past, no correlations of the infrared TF residuals with inclination, size, 
concentration, gas fraction, or far infrared luminosity have thus far been 
reported (Aaronson \& Mould 1983; V01).  The dependence of near-IR TFR
scatter on color and surface brightness is however still a matter of 
contention that we discuss in \se{color}.  The SLR scatter is
also addressed in \se{color}.

The study of scaling relations in galaxies has benefited from the 
two-pronged application of the TFR for the purposes of: (i) Estimating 
relative distances to measure deviations from  the mean Hubble flow (see, 
e.g., Strauss \& Willick 1995 and the reviews in the ``Cosmic Flows 1999'' 
proceedings by Courteau, Strauss, \&  Willick 1999); and (ii)  testing 
galaxy formation and evolution models (Dalcanton, Spergel 
\& Summers 1997; MMW98; vdB00). 
The  philosophy of sample selection and calibration differs  in both
cases. For cosmic flow analyses, the calibration and science samples
must be pruned mostly on the basis of morphology and visual appearance
in order to minimize  systematic errors and  thus ensure the smallest 
possible magnitude (distance) error. Since TF scatter depends strongly 
on the slope of the TF, it is found that the combination of steepness,
magnitude errors, extinction correction, and sky stability favors red
($R \& I$) bands for smallest distance errors and cosmic flows applications 
(Courteau 1997, hereafter C97; V01).  The accuracy
of bulk flow solutions also depends on the size of the sample. In
order to collect large enough samples, TF calibrations for flow studies 
rely mostly on \hi line widths or \ha rotation curves that can
be collected relatively quickly on modest aperture telescopes.  These
rotation measures typically sample the disk rotation out to 2 to 3
disk scale lengths (C97).  

By contrast, use of the TFR as a test bed for galaxy 
formation models requires the widest range of morphological types, 
to sample all structural properties, and that extinction and stellar
population effects be minimized to isolate genuine dynamical
correlations. The nearly dust-insensitive $K$-band is thus the one  
of choice for such applications (V01)\footnote{We find in \se{IRTF}
a slightly tighter 2MASS TFR at $J$-band rather than at $K$; the $H$-band TFR 
is hardest to control due to airglow fluctuations (Jarrett \etal 2003).}.  
Rotation velocities are
preferably extracted from fully resolved \hi rotation curves
obtained using aperture synthesis maps that sample the
disk rotation out to 4 to 5 disk scale lengths.

Ideally, the study of scaling relations should rely on homogeneous
samples assembled with the very purpose of testing for broad structural
and dynamical differences; at the moment, we must contend with the more 
finely pruned heterogeneous samples of late-type spirals that have been 
collected during the last decade mostly for flow studies.
These data, which include near-infrared luminosities, and sometimes 
colors, for large samples of galaxies, still enable us to characterize 
the dependence of scaling relations and examine various constraints of 
structure formation models, in ways hitherto unsettled.  

In Courteau \etal (2003), we showed that barred and unbarred galaxies 
have similar physical properties and that they share the same TFR. 
As a natural extension of this study and CR99, in \se{color} we use
the extensive all-sky distance-redshift catalogs presented in
\se{data} to characterize the scatter of the TF and SL relations in
terms of galaxy observables for a broad suite of galaxy types. 
In \se{CR03} we bolster the notions of
surface brightness independence and color dependence of the TFR, as
well as that of surface brightness dependence for the SLR, for disk
galaxies.  This prompts a renewed examination of the correlated
scatter study of CR99 which favored high dark-to-luminous mass
fractions in galaxy disk's interiors.

In \se{origin} we develop a simple analytical interpretation for the 
origin of the galaxy virial relations, such as the TF and SL relations, 
in a cosmological setting.  The analysis of the scatter in the scaling 
relations, and a discussion about the implications of our results in terms 
of existing galaxy formation models, are presented in our companion paper 
(S. Courteau \etal 2004a, in preparation; hereafter Paper II). 

\section{Available Samples}{\label{sec:data}}

We consider four samples for which accurate galaxy observables, including
crucial rotational velocities, are
available. These include (a) the large $I$-band survey of galaxy  
distances for bright field spirals in the Southern sky by 
Mathewson,  Ford, \& Buchhorn (1992; hereafter ``{\bf MAT}''); 
the all-sky $I$-band TF surveys by (b) Dale \etal (1999; hereafter
``{\bf SCII}'') and (c) Courteau \etal~(2000; 2004b, in preparation; 
hereafter ``{\bf Shellflow}''); and (d) the multi-band $BVRK$ TF survey 
of Ursa Major cluster galaxies by Tully \etal (1996) and V01 (hereafter 
``{\bf UMa}''). 
The first three TF surveys were originally designed to map the convergence 
of the velocity field  on $\sim 60h^{-1}$ Mpc scales, with the SCII and 
Shellflow studies paying special attention to TF calibration errors 
between different observatories.  By design, these surveys favor late-type 
galaxies with inclinations on the sky greater than 32\deg; most have 
$i\simeq 60$\deg. 

Properties for each sample are given in Table 1.  These include in 
Col.~(2), the number of galaxies from the original database that have 
a full complement of useful observables; Col.~(3), the nature of the 
sampled galaxies (cluster or field; the predominantly 
``field'' surveys of MAT and Shellflow include a small fraction of cluster 
galaxies); Col.~(4), the digital photometric coverage. 
$B$-band magnitudes for all the MAT galaxies were extracted from the RC3
(de Vaucouleurs \etal 1991).  We have 
{\it Sloan Digital Sky Survey} (York \etal 2000, hereafter SDSS)
 $g-i$ colors for 39 SCII galaxies (see Appendix B) and
 $JHK$ Kron magnitudes and colors from the {\it Two Micron All-Sky Survey} 
(Skrutskie \etal 1997; hereafter 2MASS) for the brightest $\sim 400$ SCII 
galaxies; Col.~(5) the magnitude or diameter limits of the original catalog; 
Col.~(6), the redshift limits of the survey in \kms.  For UMa, we list 
the adopted cluster distance in Mpc; and in Col.~(7),
the rotation measure, either extracted from \hi line mapping or \ha 
spatially resolved rotation curves.  While MAT gives 
both \hi and \ha rotation measures we use only the sub-sample of galaxies
with available {\hi}, which is largest.  Our results for the MAT sample do
not depend upon this particular choice of rotation measures. 
For SCII, rotation velocities were obtained using 
either \ha long-slit spectra  or \hi line profiles. These two rotation 
measures are comparable as long as the rotation curve is flat beyond 
2 disk scale lengths and the \hi surface density drops rapidly in the 
outer disk (C97).  
Rotation velocities for the Shellflow sample were measured at
$R\simeq 2.2\Rd$ from resolved \ha rotation curves. 
For the UMa database, we use the ``\hi'' sample of 44 galaxies 
with \hi line widths (V01).  

Disk scale length measurements were computed in somewhat different ways
for each sample.  For the MAT galaxies, we have used the 
two-dimensional bulge-disk (B/D) decompositions of MAT $I$-band images 
by Byun \& Freeman (1995). These decompositions assumed a de~Vaucouleurs bulge 
profile which is not ideal for late-type galaxies (MCH03) and therefore 
the disk central surface brightnesses are likely biased low. 
The disk scale lengths are less affected, in a relative sense due
to the broader baseline of the disk modeled as an exponential 
profile, and deemed adequate for the purpose of our study (see \se{virial}).  
SCII disk scale lengths were obtained by fitting a straight line to 
the exponential part of the $I$-band surface brightness profile from 
${\sim}21$ $I$-\magarc  to ${\sim}25$  $I$-\magarc (the so-called ``marking 
the disk'' technique) and corrected for projection effects according 
to $\Rx  = \Rx^\circ [1 + 0.4\log(a/b)]$,  where $a$ and $b$ are the 
semi-major  and semi-minor axes of the  disk (Dale \etal 1999; G97).  
The correction for projection is small 
and somewhat uncertain but the choice of observed or deprojected scale 
lengths for the SCII galaxies does not affect our final results.
Shellflow scale lengths were extracted from one-dimensional B/D 
decompositions of azimuthally-averaged  $I$-band surface brightness profiles 
(MCH03; S. Courteau \etal 2004; in preparation).  These fits account 
for a S\'ersic bulge and an exponential disk. 
Disk scale lengths for the UMa sample were measured from a ``marking the
disk'' technique but the fit baseline is unspecified and erratic fits 
are reported in Tully \etal (1996).  The latter could be due to inclusion 
of bulge light in the disk fit (scale lengths biased low) for the brighter 
galaxies and sky domination (scale lengths biased high) for the fainter 
galaxies.  The latter signature is detected when we compare the scale 
length measurements for the four samples in \se{virial}. 

Physical parameters are computed using \hub=72 \hunit.  This matches 
the revised distance to the UMa cluster from 15.5 Mpc to 18.6 Mpc, based 
on new Cepheid distances from the HST Key Project (Tully \& Pierce 2000).
The exact choice of distance scale does not affect our conclusions
so long as it is consistent for all samples in order to put absolute 
scales on equal footing (luminosities in $\lsol$ and scale lengths in kpc). 

Hubble types for all but the SCII galaxies were obtained from the 
heterogeneous NASA Extragalactic Data (NED) Base.  Most SCII galaxies 
were not classified in NED and their morphological types were determined
in a homogeneous fashion via a combination of eye-ball examination, 
B/D ratio and/or concentration index (Dale \etal 1999). 
Interacting and disturbed galaxies were rejected in all samples. 
Together, the MAT,  SCII, Shellflow, and UMa samples combine
for a total of 1,750 separate entries; some repeat measurements 
within a given sample exist but all observations are independent. 

\subsection{Mean Parameter Relations}{\label{sec:virial}}

\begin{figure}
\plotone{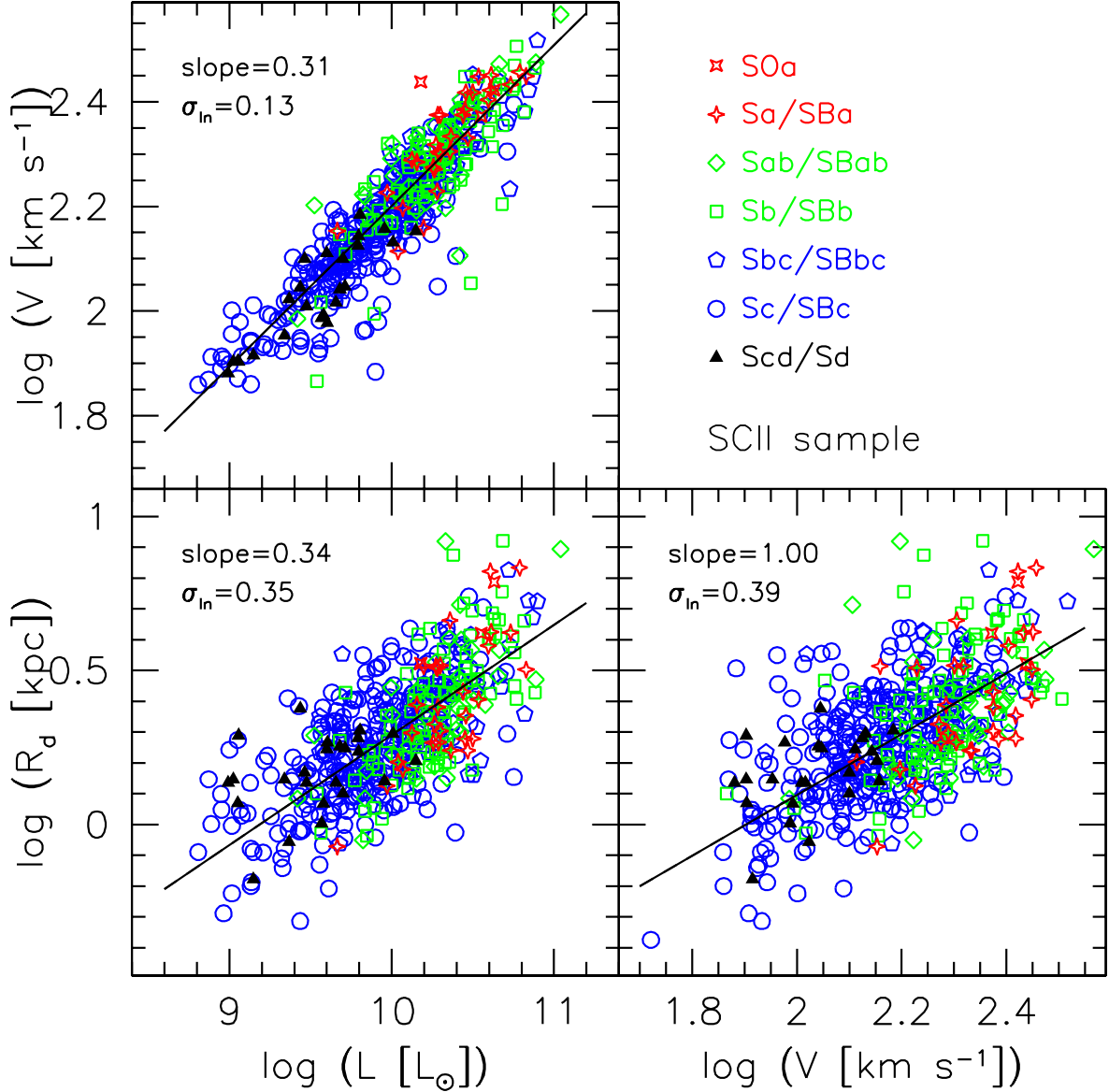}
\caption{Scaling relations for SCII galaxies.  Line widths are measured 
 from \ha rotation curves and \hi line widths and disk scale lengths are 
 measured using the ``marking the disk'' technique (see text).  
 Luminosities are computed from fully-corrected $I$-band magnitudes. 
 Unlike Dale \etal the magnitudes do not include any correction 
 for morphological type dependence. The solid lines correspond to the 
 linear regressions to the SCII galaxies, based on our data$-$model 
 minimization technique.  The inset gives the slope of the fit and
 the scatter of the ln-normal distribution.
\label{fig:SCII_3rel}}
\end{figure}

We consider the projected distribution in each of the planes defined
by a pair of the three log virial variables, $\log \Rd$, $\log \Vobs$, 
and $\log L$; these are shown in Figs.~\ref{fig:SCII_3rel}--\ref{fig:UMa_3rel}
for the SCII, MAT, Shellflow, and UMa galaxies. The axis limits are
the same in all figures for trend matching between each sample. 
Different symbols identify the full range of spiral Hubble types, 
as a proxy for accurate optical colors which are not available 
for the full MAT and SCII samples.  The solid lines 
correspond to one dimensional linear regressions to (a)
$\log\Vobs$ on $\log L_I$, (b) $\log\Rx$ on $\log L_I$, and (c) 
$\log\Rx$ on $\log\Vobs$, given the measurement errors in the first 
variable in each case when available (or otherwise using unweighted 
fits). The fits for the SCII sample are used as fiducials 
against the other, less complete, samples and are shown as a 
dotted line on the virial projections for each sample.  
For each relation, we fit the
functional form $y(x) = \bar{y}  + \alpha_y  (x - \bar{x})$, 
where $y$ and $x$ stand for any combination of $\log\Vobs$,
$\log\Rx$, and $\log L$; the upper bar denotes the median value 
of each variable. To achieve a robust fit to the
data $\bigl [y(i),x(i)\bigr ]$, we varied $\alpha_y$ 
to minimize
the data$-$model  deviation (rather than the  squared difference). We
repeat the fits for all the log quantities keeping either the 
total luminosity, stellar scale length, or maximum velocity fixed. This
fitting technique follows the treament in CR99.  We have also computed 
standard least-squares fits which yield similar results.  
Log-slopes and the Pearson linear correlation coefficients $r$ for 
each parameter combination are reported in Table~2 (we do not concern
ourselves with calibration zero-point differences in this Paper). 

The regressions were performed using the full range of luminosities
(when the regression was on $\log  L$) and velocities (when the
regression was on $\log \Vobs$).  It has been proposed that the 
scaling relations may be different for the fainter, LSB galaxies 
(e.g., Kauffmann \etal  2003; Shen  \etal  2003), where supernova 
feedback effects are likely to be important (e.g., Dekel \& Silk 1986;  
Dekel \& Woo 2003).  Such a departure is, however, not apparent in 
the samples (SCII and MAT) that most closely probe the LSB regime 
(see Figs.~\ref{fig:SCII_3rel} \& \ref{fig:MAT_3rel}). 

The agreement in Table 2 between the log-slopes for different samples 
is relatively good, though differences exist.  All samples are well 
matched by a TFR, $V \propto L^\alpha$, with log-slope $\alpha_I=0.31$.  
Our TF fits are also a close match to those published by the original 
authors (as $M_{\rm abs} \propto \log \Vobs$), even though 
the minimization techniques can be quite different.
The Pearson correlation coefficient of the distribution (not
considering the errors) averages $r=0.90$ for the TFR and the conditional
distribution of $\Vobs$ at a given $L$ is roughly log-normal. The
correlations for the SL and size-velocity (SV) relations are weaker, averaging
$r\sim0.6$. The size distributions are less robust than the TFR 
owing to a number of factors such as, (a) small range of galaxy sizes 
in some samples (Shellflow, UMa), (b) lack of a uniform and universal 
definition of scale length (MCH03), and (c) intrinsic scatter due
to the natural dispersion of the spin parameter, $\lambda$ (\se{origin}).
Whether scale lengths are measured from B/D decompositions or ``marked'' 
over a specified range of surface  
brightnesses (while omitting the bulge region or not) can yield scale 
differences greater than 20\% (MCH03). 
The overlap between the four samples is too small (in some cases 
non-existant) to calibrate such systematic  errors.  Thus it is difficult
to determine how
much of the lower correlation coefficients in the SL and SV relations
is due to genuine dynamical processes and$/$or inadequate scale lengths.
In spite of potential pathologies with the MAT scale lengths (see \se{data}), 
the scalings for this sample and SCII are comfortably close.  For the sake 
of uniformity, we will base our final analysis of the virial relations 
of disk galaxies on the SCII sample alone, being the most 
extensive and robust sample considered here.  Furthermore, unlike our
results for some of the other, less complete and thus noisier, data sets
our fits for SCII are self-consistent, within the errors. 
If $V \propto L^\alpha$  and  $R\propto L^\beta$ and $V \propto R^\gamma$,
then self-consistency requires that $\alpha = \beta \gamma$.
This is indeed the case for SCII, our most complete sample.

In summary, we find the following mean power-law scaling relations 
at $I$-band for spiral galaxies: 
\be  
\Vobs \prop L_I^{0.31}, \quad  \Rx \prop L_I^{0.34}, 
\quad  \Rx \prop \Vobs^{1.00}.
\label{eq:scaling}
\ee
The uncertainties in the log-slopes for the SCII sample, based 
on measurement errors, are $\pm 0.02$, $\pm 0.02$, and $\pm 0.09$,
respectively.

\Fig{allTF} shows the combined TF and SL relations for the four
samples.  This operation formally requires that the magnitudes, 
rotation velocities, and scale lengths be perfectly homogeneous.  
However, zero-pointing differences exist between the various 
samples thus preventing a direct merging of all samples into one 
large homogeneous catalog for our study of scaling relations.  
The rest of our investigation thus relies on the separate
examination of each data set. 

\subsection{The Morphological/Color Dependence of Galaxy Scaling 
 Relations}{\label{sec:color}}

The morphological type dependence of the TFR is highlighted in 
Figs.~\ref{fig:SCII_3rel}--\ref{fig:UMa_3rel} by different 
point types and colors.  This dependence, first noted by Roberts 
(1978), and revisited by Aaronson \& Mould (1983), Rubin \etal (1985), 
Giraud (1986), Pierce \& Tully (1988) and G97 
-- to name a few -- goes in the general sense of early-type 
galaxies rotating faster than later-types at a given optical 
luminosity.  This trend is confirmed for each sample in the top 
panels of Figs.~\ref{fig:SCII_3rel}--\ref{fig:UMa_3rel} where 
the early-type and late-type spirals, represented by red stellated 
points and black filled  triangles, are located above and below the 
fiducial fits, respectively\footnote{At the lenticular end of the
spiral sequence, S0 galaxies are consistent with our qualitative
findings with a marked offset from the mean TFR 
(of Pierce \& Tully 1992) but these galaxies also show a steeper 
TFR than early-type spirals (Neistein \etal
1999; Hinz, Rix, \& Bernstein 2001; Mathieu, Merrifield, \& Kuijken 2002).
The steeper TFRs may be a consequence of sample selection and$/$or 
just the nature of S0 galaxies.}.  Recent optical TF relations of 
nearby spirals (\eg KFF02) and of distant spirals (B. Weiner 
\etal 2003 [DEEP project], in  preparation) covering a wide range of 
morphologies support the Hubble-type  dependence of the optical
TFR, up to at least $z \sim 1$.  

Figs.~\ref{fig:TFres_shell} \& \ref{fig:TFres_SCII} show histograms
of the luminosity residuals from the mean TF relation as a function 
of Hubble type for the Shellflow and SCII samples (similar results are 
obtained for MAT and UMa). By selection, 
the mean TF fit is dominated by Sb-Sc galaxies. For the Shellflow sample, 
early and late-type spirals differ by 0.08 dex and -0.25 dex from the 
mean TFR respectively. This is accentuated in the SCII sample 
(Fig.~\ref{fig:TFres_SCII}) with TF
luminosity residuals of +0.17 dex and -0.33 dex for early and
late-types, respectively. Dale \etal (1999) applied a
morphological-type correction to their galaxy magnitudes for the 
construction of their TF template.  Their corrections to total $I$-band
magnitudes for T$\leq$Sab, Sb, and T$\geq$Sbc are -0.27, -0.11, and 0.00 
mag, respectively, which is somewhat larger than the ones we find for 
their data.

While Hubble-type classification depends on the bandpass of selection, 
a morphological-type dependence of the TFR, even at near-infrared bands,
is undeniable.
The morphology dependence of the SV and SL relations is weaker, as one
would expect based on the relatively larger errors of disk scale lengths 
compared to circular velocity measurements.  The distributions of scale
lengths in the SL diagrams show a slight excess for bluer (later type) 
galaxies over the mean at a given luminosity, suggesting that bluer 
galaxies are bigger than the mean. 
Prejudice from theoretical models (e.g. FAR00; vdB00) 
and our own simple scenario of disk formation (\se{origin}) would have us 
favor a scenario where redder objects are both faster rotators and more 
compact, but observational evidence for the latter is weak.  This could
be the result of inhomogeneous disk scale length measurements, and 
especially for the fainter systems for which sky errors are more conspicuous.
The slight excess of disk scale lengths over the mean at the faint end of 
the SLR could be understood if sky subraction was systematically 
under-estimated.  In that respect alone, an extensive program to measure 
accurate disk scale lengths and half-light radii for a large collection 
of spiral galaxies based on one unique reduction method and high 
signal-to-noise data is badly needed. 

Because morphological classification is a subjective 
measure, we look at more objective observables to determine what
physical parameters drive the TFR scatter.  Figs.~\ref{fig:TFmag1} 
and \ref{fig:TFmag2} display the TFR scatter in magnitude against 
five galaxy observables for the four samples.  
Besides the RC3 and SDSS colors extracted for the MAT and SCII samples
respectively (see Appendix B for SCII), the data were all collected 
from the original literature. 
The concentration indices (CIs) for Shellflow and SCII galaxies correspond 
to the ratio of radii containing 75 and 25 percent of the total light (CI72).
CIs for UMa galaxies were computed at radii containing 80 and 20 percent 
of total light (CI82; Tully  \etal 1996).  Note that all exponential disks 
without a bulge have the same value of CI72=2.8 and CI82=3.6, regardless 
of their total mass or scale length. For the MAT sample, we substituted 
missing CI data with B/D ratios.  

Looking at Figs.~\ref{fig:TFmag1} \& \ref{fig:TFmag2}, we see that the 
TFR scatter (in magnitudes) is clearly independent of central 
surface brightness, $\mus$, physical disk scale length, $\Rd$,
and concentration index. Color and morphological types, on the other
hand, are both correlated with TF magnitude residuals. Given that
color and morphological type are themselves strongly correlated (see
\Fig{Hubtype}), this is consistent with a picture in which color is
the fundamental driver of scatter in the TF relation.  \Fig{Hubtype}
shows that the tightest correlation with morphological type is indeed
with color.  \Fig{Hubtype} also shows that velocity depends more
strongly upon morphological type (color) than disk size, an important
observation that we return to in \se{CR03} and Paper II.  The correlation 
of morphological types with concentration index is poorest, highlighting
the relative inadequacy of the CI test to discriminate between
different spiral Hubble types.

According to V01, all
correlations of TF residuals with galaxy observables detected in blue
bands, including color, vanish in the infrared.  This important conclusion 
however 
rests on the study of a small data sample (less than 44 galaxies) 
and deserves closer attention.  Contrary  to findings in Aaronson \& 
Mould (1983) and V01, Rubin \etal (1985) found that the morphological 
type dependence of the TFR was reduced, but not eliminated, at infrared 
$H$ luminosities\footnote{
Unlike V01 and Aaronson \& Mould (1983), Rubin \etal (1985) found 
a similar morphology (color) dependence of the TFR, even at $H$-band.  
Such a dependence could be an artifact of rotational 
speed measurements with optical spectra.  The shape of optical 
rotation curves is often dominated by the inner rising part of 
an early-type spiral or the slowly rising part of a later-type 
spiral of the same luminosity.  However, the presence or absence of 
a bulge or bar has a weaker effect on the overall extended shape of 
a resolved \hi rotation curve.  Thus the asymptotic rotational speed
measured from \hi synthesis maps or \ha velocity fields from say, 
2 disk scale lengths to 
the last detected velocity point (with acceptable signal-to-noise), 
will not be subject to the vagaries of the inner rotation curve in the 
same way as optical rotation curve measurements (\eg C97, V01).  
So long as velocity measurements are made beyond $R=2.2 \Rd$,
the rotation speed should be unaffected by a bulge or bar 
(CR99; Courteau \etal 2003), and the trends detected by Rubin 
\etal and ourselves are robust.}.  Thus color is potentially another
fundamental parameter of galaxy structure and formation, in addition 
to size, luminosity and rotation speed.  Until recently, the 
CCD luminosity measurements in large galaxy samples with available 
line widths were only collected in a single band (typically $r$ or $I$), 
thus thwarting any systematic test for color dependence.  Recent 
scatter studies based on the small, pruned, UMa galaxy sample 
(Heavens \& Jimenez 1999; V01) can now be duplicated for much 
larger samples, as we do here. 

The recent $R$-band TF study of the Nearby Field Galaxy Survey (Jansen
\etal 2000) by KFF02 identified $B-R$ color and \ha equivalent width
as the main drivers of the TFR scatter.  Color and \ha equivalent
width are both tributary of star formation histories though the former
depends both on the {\it integrated} and instantaneous star formation
while the latter is a function of the current star formation rate
alone. Taken together with our results, it thus seems that star
formation history is the fundamental driver of the TF scatter (e.g.
Heavens \& Jimenez 1999). One expects the TF scatter in
bluer bands to be more sensitive to contributions of 
instantaneous star formation activity, while at (near-)infrared bands
the scatter mainly reflects the convolved star formation histories.  

In order to verify that the basic trends in the scatter of the galaxy
virial (TF and SL) relations are a manifestation of star formation 
{\it histories}, and thus persist at infrared bands, we need to expand 
our data base as we do below. 

\subsection{The Infrared Tully-Fisher and Size-Luminosity 
            Relations}{\label{sec:IRTF}}

The advent of large-scale infrared surveys such as 2MASS provides 
us with reliable $JHK$ luminosities, effective radii, and colors to 
construct (nearly) dust-free scaling relations.  With a typical surface 
brightness limit of $K \sim 20$ mag arcsec$^{-2}$, 
the 2MASS luminosity profiles are a full two magnitudes 
shallower than the typical $I$-band profiles in our samples 
or the SDSS brightness profiles.  Yet, they yield scaling 
relations that are as tight as the ones derived 
at $I$-band.  We now restrict our IR-extended analysis to 
the SCII sample for which 
the greatest range of luminosities and rotation speeds is found and 
thus for which the most reliable scaling relations can be derived.  
Because of the 2MASS magnitude limit of 13.5 $K$-mag (e.g. 
Bell \etal 2003b), only the $\sim{400}$ 
brightest SCII galaxies have measured Kron magnitudes. 

Figs.~\ref{fig:2MASS_Reff} to \ref{fig:2MASS_JmK} show the 2MASS
$J$-band scaling relations for SCII galaxies plotted against $\Re$,
$\SBe$, and $J-K$ colors\footnote{Concentration indices from the 
2MASS data base are known to be pathological (Bell \etal 2003b),
as we verified ourselves.  Consequently, we omit scaling relations 
with CI as a discriminant.}. In the $H$-band one is notoriously more
sensitive to airglow fluctuations (e.g., Jarrett \etal 2003), making
this photometric band less attractive for TF studies. The $K$-band TFR
is very similar to that in the $J$-band, but since the latter yielded
the smallest TF scatter, we focus on the $J$-band in what
follows. The scatter in the SL and SV relations is significantly
reduced compared to that in the $I$-band (cf.~Table 2).
Note, however, that the size parameter in the $J$-band is an effective
radius rather than a disk scale length (which was not available from
the 2MASS data release). Therefore, it is unclear whether this
reduction in scatter reflects the usage of a more robust and rigorous
galaxy size measure, or whether it is related to stellar population
differences.

As reported in Table 2, the (SCII) $I$ and $J$ TFRs have nearly the
same log-slopes.  For the SCII sample with 2MASS data, the infrared
scaling relations are:
\be  
\Vobs \prop  L_J^{0.30}, \quad \Rd \prop  L_J^{0.41}, \quad 
\Rd \prop  \Vobs^{1.36}.
\label{eq:IRscaling}
\ee

While the log-slopes for our SL and SV relations in 
Figs.~\ref{fig:2MASS_Reff} to \ref{fig:2MASS_JmK} were computed with 
respect to $\Re$, they can be compared directly to $\Rd$ since,
for a pure exponential disk, $\Re = 1.678 \Rd$.  However, the $J$-band
SL and SV log-slopes are significantly steeper than at $I$, possibly 
resulting from biases in the measurement of disk scale lengths versus
effective radii.  The former is sensitive to the disk fitting baseline 
while the other depends on the strength of the bulge-to-disk ratio. 
In order to verify this, we would need both types of radii measurements
(either at $I$-band or in the infrared) which are not currently available; 
the assembly of such a data base is under way. 
Low and intermediate-redshift ($z\sim1$) studies of SV
relations, based on data from Courteau (1996; C97) and Vogt \etal
(1997) and as reported in Mao, Mo, \& White (1998), have SV log-slopes 
of $1.07$ and $0.96$ respectively, in good agreement with our $I$-band
measurements.  Dynamical considerations (\se{origin}) also suggest a 
log-slope of unity for the SV relation.  More work 
needs to be done to ascertain the cause of the steep infrared 
size-dependent scaling relations. 

Figs.~\ref{fig:2MASS_Reff}--\ref{fig:2MASS_JmK} confirm
that the TF scatter is dominated by color and nearly independent
of size and surface brightness.
Previous reports of TF residual correlations with surface 
brightness may have generated confusion.  Willick (1999) reported 
a correlation of $I$-band TF residuals with surface brightness and 
compactness for his ``LP10K'' survey of distant cluster galaxies 
using a moments fitting method to determine the exponential disk 
parameters even in the presence of irregularities in the galaxy 
light profiles.   In an attempt to alleviate subjective fitting 
boundaries, Willick's computation used the entire surface brightness 
profile, including the bulge, and thereby biasing scale length 
and central surface brightness measurements.  Applying his procedure, 
we can reproduce the putative surface brightness dependence of the TFR 
while proper bulge-to-disk (B/D) fitting techniques (MCH03) find 
none\footnote{It is unfortunate that J. Willick
is no longer with us to defend his approach.}. 

The absence of 
surface brightness dependence of the TFR scatter has been verified by 
Sprayberry \etal (1995), Zwaan \etal (1995), and V01, and suggests
a near-constancy of near-infrared (NIR) $M_d/L$ 
ratio for high surface brightness (HSB) and low surface brightness (LSB) 
galaxies of comparable total luminosities.  In other words, if the 
surface brightness decreases, so do $\Vobs$ and $L$ in such a way
that the TFR is independent of surface brightness. 
Thus, HSB and LSB galaxies belong to the same TFR, albeit with 
LSB galaxies showing a greater spread at lower luminosities.  
Any TF offset between low and high surface brightness noticed in the 
$B$-band can be explained by mass-to-light ratio differences between 
the relevant stellar populations and, according to V01, the offset
disappears at $K$.  

The SLR shows a very strong dependence on $\Sigma_{\rm eff}$
(as expected since $\Sigma_{\rm eff} \propto L/\Rd^2$), and a weaker 
dependence on color.  We see from \Fig{2MASS_JmK} that bluer galaxies 
are on average larger than the mean, at a given luminosity, but we
cannot state with any confidence that redder galaxies have smaller
characteristic sizes, since these scatter evenly about the mean line. 

\section{Residual Correlations of Scaling Relations}{\label{sec:CR03}}

The previous section identified color and surface brightness as the 
primary source of scatter in the TF and SL relations, respectively.  
In the spirit of CR99, we now examine correlations of the residuals 
from the mean virial relations in each sample.

Following \se{virial}, we define the residuals for each object $i$
as $\Delta y(i)\equiv y(i)-y_{fit}(i)$.  Fig.~\ref{fig:3res} shows 
residual correlations for combinations of $\dlogV$, $\dlogR$, and 
$\dlogL$ for each sample based on the $I$-band catalogs. The colored 
types have the same morphological dependence as in 
Figs.~\ref{fig:SCII_3rel}--\ref{fig:UMa_3rel}.  
The correlation residuals for the SCII sample with 2MASS $J$-band 
data are presented in \Fig{2MASS_3res}; the point types are
a function of $J-K$ color. 
As in CR99, we apply a robust non-parametric test by rotating the set
of residuals $\bigl (\Delta y_1(i),  \Delta y_2(i) \bigr )$ by various
angles   $\theta$   to  get   $\bigl   (  \Delta   \hat{y}_1(i),\Delta
\hat{y}_2(i) \bigr  )$ and then  applying a Spearman rank  test (Press
\etal 1992,  \S 14.6) for  correlations between the  quantities $\bigl
(\Delta  \hat{y}_1(i), \Delta  \hat{y}_2(i)  \bigr)$.  The  acceptable
range  of  $\partial  (\Delta   y_1)/\partial  (\Delta  y_2)$  can  be
calculated  from the  range of  angles $\theta$  for which  the $\bigl
(\Delta\hat{y}_1(i),\Delta\hat{y}_2(i)   \bigr   )$   are  {\it   not}
significantly correlated.  The correlation slopes and their associated
errors are  reported in Table 3 and are shown at the bottom of each
panel in \Fig{3res}. The linear correlation coefficients, $r$, which  
provides another means of assessing the strength of the
correlation between each residual, are also shown at the top left 
corner of each panel.  The slopes and correlation coefficients shown
at the bottom of \Fig{2MASS_3res} are representative of the full data
set (including all colors). 

There is close agreement in the correlated residual solutions of 
all samples, and including the 2MASS SCII data set, with the general
solution: 
\begin{eqnarray*} 
  \partialVRL & = & -0.07 \pm 0.05  \\
  \partialRLV & = & \phantom{-}0.12 \pm 0.03 \\
  \partialVLR & = & \phantom{-}0.29 \pm 0.02.
  \label{eq:CR03}
\end{eqnarray*}
These slopes are an average of the values for Shellflow, MAT, and the
$I$ and $J$-band SCII samples.  Besides UMa, the slopes for the TF/SL
residuals for all samples are negative (anti-correlated) and low.  For
the three large samples and at all colors, the TF/SL residual
correlation is statistically different from $\partialvr=0$, though
only weakly.  Furthermore, redder galaxies (top left corner of
\Fig{2MASS_3res}) lie above the null line whereas blue galaxies
(bottom left) lie below.  Note also that galaxies of all morphological 
types (and barredness, not shown here), scatter normally about the zero 
line. FAR00 have also computed the TF/SL residuals for the UMa sample
(V01; see lower left panel of \Fig{dvdr}) and find no correlation 
for both for LSB and HSB galaxies, as do we for that sample (see Table 3).
That result is however heavily weighted by the small size of the sample.

The weak and strong surface brightness dependences of the TF and SL
relations, respectively, have been used to infer the ratio of baryonic
to dark mass in the inner parts of galaxies (CR99).  If the gradient
of NIR mass-to-light ratios is self-similar among galaxies of
different brightnesses, CR99 conjectured that maximal disks\footnote{A
  stellar disk is ``maximal'' if it contributes more than 75\% of the
  total rotational support of the galaxy at $R\equiv 2.2\Rd$, the
  radius of maximum disk circular speed (Sackett 1997). $\Rd$ is the
  scale length of the disk.} should be offset from the mean TF and SL
relations in such a way that the fractional deviations,
$\Delta\logV(L)$ and $\Delta\log R(L)$, from the mean relations,
$V(L)$ and $R(L)$, should be strongly correlated. The case of
$\partialvr=-0.5$ would be expected for pure (maximal) exponential
stellar disks (i.e., without a dark matter halo).  Uncorrelated TF/SL
residuals might then suggest that disk galaxies (of all colors, types,
and barredness) are, on average, sub-maximal\footnote{The most massive
  spiral galaxies with $V_{max}>200$ \kms\ may harbor disks with high
  mass fractions, in agreement with the maximum disk hypothesis
  (Athanassoula, Bosma, \& Papaioannou 1987; Kranz, Slyz, \& Rix
  2003).} and$/$or that a fine-tuning of halo and disk parameters 
washes out any surface brightness correlation in the TFR.  Using the
samples of bright late-type spiral galaxies of Courteau (1992) and
MAT, with available $R$ and $I$-band luminosity profile and \ha
rotation measures, CR99 found a tentative correlation between the TF
residuals (measured in velocity space) and the SLR residuals.  The
TF/SL fractional deviations were found to be weakly correlated with
$\partialVRL=-0.19 \pm 0.05$.  With our expanded study, we now find
$\partialVRL = -0.07 \pm 0.05$, suggesting that the disk mass plays
only a small role in setting the total velocity at $2.2\Rd$.
According to the galaxy contraction models of CR99\footnote{Our models
  have been independently confirmed by Bullock (2002; priv. comm.)},
the residual correlation \break $\partialVRL \lta -0.07 \pm 0.05$
corresponds to $V_{\rm disk}/V_{\rm total} \leq 0.55 \pm 0.10$,
indicating that spiral galaxies, independent of barredness or type,
would be dominated by dark matter with $M_{\rm dark}/M_{\rm total}
\geq 0.70$ within $2.2\Rx$.  This result agrees with analytical models
of cosmological structure formation (\eg Dalcanton \etal 1997; MMW98;
NS00; vdB00; Cole \etal 2000), whereby dark halos dominate the
potential even in the inner parts of galaxies, and with independent
measurements of galaxy dynamics (e.g.  Fuchs 2001; Trott \& Webster
2002; Kranz, Slyz, \& Rix 2003).  This result does not preclude the
possibility of a stronger disk at radii smaller than $2.2\Rx$.

With $\partialRLV=0.12 \pm 0.03$, the SL residuals (middle column of 
Figs.~\ref{fig:3res}--\ref{fig:2MASS_3res}) are also weakly 
correlated.  A positive correlation may be real but not of fundamental
interest.  Consider a model where V(R) versus L(R) is a strict relation 
with no scatter, while R is a scattered quantity about the relation R(L).
Then the scatter in both V(R) and L(R) comes solely from the scatter in
R, and of course the residuals will be correlated, both determined by 
the value of R for a specific galaxy.  Thus it is the TFR between 
V and L that gives rise to the V(R)-L(R) correlation.  In a similar way, 
the weak SL correlation between R and L gives rise to a (weak) correlation 
between the residuals of R(V) and L(V).  Models to explain the numerical 
value of the correlation residuals are presented in Paper II. 

The VL residual distributions (third column of 
Figs.~\ref{fig:3res}--\ref{fig:2MASS_3res})
are tightly correlated with  $r\sim0.75$ (a little less at $J$) and relatively 
small slope errors; indeed, this relation is simply the TFR recast in 
differential form. 

Figs.~\ref{fig:dvdr} and \ref{fig:2MASSdvdr}
are expanded versions of the $\partialVRL$ relations for 
the MAT, SCII, Shellflow, and UMa samples with $I$-band data, 
and the SCII sample with $J$-band data, respectively. 
Again, the correlated residuals are fully inconsistent with the prediction 
for a pure stellar exponential disk (short-dashed line), 
$\partialVRL=-0.5$ at $\Rx$ (CR99).  The solid and long-dashed lines 
show the best fits from our Spearman rank test.
The lower panels in Figs.~\ref{fig:dvdr} and \ref{fig:2MASSdvdr} 
show TF and SL residuals with color.  Trends of $\dlogV$ and 
$\dlogR$ with color residuals are shown in the right lower and central
panels; the residual fits for the SCII+2MASS data are shown as solid 
and long-dashed lines in \Fig{2MASSdvdr}.  
Velocity residuals correlate tightly with color whereas size residuals 
are noisier and statistically consistent with zero. 
These trends match those seen in CR99 for the Courteau-Faber and 
MAT samples and ought to be reproduced by evolutionary models of
disk galaxies (e.g. FAR00; vdB00).  If we 
account for the observed trends with color, and the simplest 
assumptions about single stellar population models, we find that 
the mean observed value of $\partialVRL \lta -0.05 \pm 0.05$ yields 
an {\it upper} limit to the contribution of disk stars to the rotation 
curve at $R=2.2\Rd$ (CR99).

\section{On the Origin of the Disk Galaxy Scaling Relations}\label{sec:origin}

The linear regressions of $\Vobs$ on  $L$ and $\Rd$ on $L$, mostly for
the  $I$-band results,  yield the  following  average TF,  SL, and  SV
relations:
\be
\Vobs \prop L^{0.31}, \quad
\Rd \prop L^{0.33}, \quad 
\Rd \prop \Vobs^{1.00}.
\label{eq:ave_scaling}
\ee
Within these  relations, at a  given luminosity, we found  that redder
spirals tend  to rotate  faster and show  hints of being  more compact
than bluer spirals.

In the spirit of the  spherical-collapse model (Gunn \& Gott 1972) one
can define the virial  radius of a collapsed, virialized gravitational
body as  the radius inside  of which the average density is  a factor
$\Delta$ times  the average background density (the  value of $\Delta$
depends on redshift and cosmology;  see e.g., Eke, Cole \& Frenk 1996;
Bryan \& Norman  1998).  It is straightforward to show (see Appendix A)
that with this definition the virial quantities $\Rv$, $\Mv$ (defined as  
the total mass within $\Rv$), and $\Vv$ (defined as the circular velocity 
at $\Rv$) are related  according to $\Mv \propto \Vv^3$  and $\Rv \propto
\Vv$.  Therefore, as long as $\Mv / L$, $\Rv / R_d$, and $\Vv / \Vobs$
are roughly constant (i.e., are not significantly correlated with
either of the virial quantities),  these virial relations are all that
is  required  to  explain  the  observed scaling  relations  for  disk
galaxies.

To realize the implications of this in terms of galaxy formation theory, 
we write
\be
\label{ratone}
\Mv / L = {\Upsilond \over \fbar \, \gf}
\ee
Here,  $\Upsilond \equiv M_d/L$  is defined  as the  disk mass-to-light
ratio, $\fbar$ is the baryonic mass fraction of the Universe ($\fbar =
\omb/\Omega_0 \sim  0.1$ for a  $\Lambda$CDM cosmology), and  $\gf$ is
the galaxy  formation efficiency that  describes what fraction  of the
baryonic  mass inside the  halo ultimately  ends up  in the  disk. The
virial relation  $\Mv \propto \Vv^3$  thus turns into the observed TFR
$L \propto \Vobs^3$ if, and only if,
\be
{\cal C}_{\rm TF} \equiv {\gf \over \Upsilond} \, \left( {\Vv \over \Vobs}
 \right)^3
\ee
is not correlated with  $\Vobs$ and reveals  an amount of scatter in
agreement with that  observed in the TF relation.  

In  the standard model  for  disk formation,  set  forth  by Fall  \&
Efstathiou  (1980), Dalcanton  \etal  (1997) and  MMW98, the  relation
between $R_d$ and $\Rv$  derives from the acquisition and conservation
of (specific) angular momentum by  the baryons, when cooling to form a
centrifugally  supported  disk.  The   total  angular  momentum  of  a
virialized system is conveniently  expressed by the dimensionless spin
parameter 
\be 
\lambda \equiv {J_{\rm vir} \over \sqrt{2} \Mv \, \Rv \, \Vv}
\ee
(Bullock \etal  2001). For an exponential disk embedded in a dark
matter halo, the total angular momentum is given by
\be
J_d = 2 \pi \int_{0}^{\Rv} \Sigma(R) V_c(R) R^2 {\rm d}R = M_d R_d \Vv f_V
\ee
with  $\Sigma(R)$  and  $V_c(R)$  the  surface  density  and  circular
velocity  of the  disk,  and  
\be
f_V = \int_{0}^{\infty} {\rm e}^{-u} \, u^2 \, {V_c(R_d u) \over \Vv}
\, {\rm d}u
\ee
(MMW98)  a  dimensionless  number.  If disk  formation  occured  under
conservation of  specific angular momentum, as  generally assumed, one
therefore obtains that
\be
\label{rattwo}
{R_d \over \Rv} = {\sqrt{2} \, \lambda \over f_V}
\ee
(MMW98). Thus,  for the virial  relation $\Rv \propto \Vv$ to turn into
the observed SV relation $R_d \propto \Vobs$, one requires that
\be
{\cal C}_{\rm SV} \equiv {\lambda \over f_V} \, \left( {\Vv \over \Vobs} \right)
\ee
is not correlated with $\Vobs$ and has a scatter around its mean value
in agreement with that in the observed SV relation.

Similarly, in order to relate the virial relation between $\Rv$ and
$\Mv$ to the observed SL relation, the quantity
\be
{\cal C}_{\rm SL} \equiv {\lambda \over f_V} \, 
\left( {\gf \over \Upsilond} \right)^{1/3}
\ee
cannot  be correlated  with galaxy luminosity,  and has to  reveal an
amount of scatter that matches that in the SL relation observed.

Thus, the challenge for galaxy formation theories is to understand how
to meet the  constraints on ${\cal C}_{\rm TF}$,  ${\cal C}_{\rm SV}$,
and  ${\cal C}_{\rm SL}$  derived here.   That this  is not  a trivial
matter becomes  clear if  one takes into  consideration that  (i) both
$\gf$ and $\Upsilon$ depend  extremely sensitively on the efficiencies
of cooling, star  formation and feedback, and are thus expected to be
strongly correlated with $\Mv$ and  $\lambda$ (see e.g., van den Bosch
2002a),  and  (ii)  that  both  $\Vobs/\Vv$  and $f_V$  depend,  in  a
convoluted way, on $\lambda$,  $\gf$, and on halo concentration, which
itself depends on halo mass (e.g., Navarro, Frenk \& White 1997).

An additional  challenge for galaxy formation models  is to understand
the dependence of  the scaling relations on disk  color. For instance,
reproducing  that color  is  the prime  driver  of scatter  in the  TF
relation, requires that ${\cal C}_{\rm  TF}$ is correlated with color. 
In hierarchical structure formation  scenarios, small mass haloes form
earlier  than their more  massive counterparts.  Since gas  cooling is
very efficient in low mass  haloes at high redshift, these models 
generally predict that fainter galaxies (living in low mass haloes)  
are older, and therefore redder, than their more luminous
counterparts. This  is opposite  to the observed  trend, and  has long
been  known to be  an important  problem for  galaxy formation  in CDM
comogonies (see Bell \etal 2003a  and MacArthur \etal 2003b for recent
discussions). Because  of this  problem, simple disk  formation models
within this CDM  framework naturally predict that the  slope of the TF
should be steeper in  bluer bands (e.g., van den Bosch 2002a), again
opposite to what has been observed.  For the more massive galaxies to
be older/redder,  one could adjust the feedback  prescription in order
to vary the amount of gas accreted by spiral galaxies as a function of
time.   However, the  energetic  requirements for  this  to occur  may
exceed the  energy reservoir in  the stellar winds and  supernovae and
would require  the action of an  AGN (Bell \etal  2003a).  The challenge 
is to realize this such that the constraints on ${\cal C}_{\rm TF}$, 
${\cal C}_{\rm SV}$, and ${\cal C}_{\rm  SL}$ are not violated.

In summary,  we can  conclude that the  observed scaling  relations of
disk galaxies are predominantly  a straightforward result of the basic
dynamics of  halo formation in  the standard cosmological  picture and
the  simple, standard scenario  of disk  formation. They  do, however,
impose some interesting constraints  on galaxy formation models, which
we address in more detail in Paper~II.

\section{Discussion}\label{sec:discussion}

By assembling some of the most extensive existing databases of 
galaxy structural parameters, and adding new infrared luminosity 
and effective radii from 2MASS, we have been able to confirm 
or infer the following major observational results: 

\begin{itemize}

\item The scaling relations of spiral galaxies, obtained by
  linear regressions of $\Vobs$ on $L$ and $\Rd$ on $L$, are 
$$
 \Vobs \prop L_I^{0.31\pm0.02}, \quad 
 \Rd \prop L_I^{0.34\pm0.02}, \quad 
 \Rd \prop \Vobs^{1.00\pm0.09}.$$

     This implies that the surface brightness 
     $\Sigma \propto L/\Rd^2 \propto \Rd \propto \Vobs \propto \Mv^{1/3}$.  
     These scaling relations seem to reflect the simplest possible model 
     for disk galaxy formation based on angular momentum conservation of
     the baryons inside virialized dark matter halos.

\item The residuals of $V(L)$ in the TFR are correlated with 
      color, and uncorrelated with effective surface brightness. 
      The residuals of $R(L)$ are correlated with surface brightness 
      and marginally correlated with color.  

\item Early-type (redder) disk galaxies rotate faster than later (bluer) 
      types at a given luminosity.  The effective radii of blue galaxies 
      are marginally larger than the mean at a given luminosity, 
      while the sizes of red galaxies do not show a significant systematic 
      deviation from the mean. 

\item The TF/SL residuals are weakly anti-correlated with 
      $\partialVRL =-0.07 \pm 0.05$. This confirms the results of 
      CR99, suggesting that most spiral disks are sub-maximal. 

\item The SV/LV residuals are weakly correlated with 
      $\partialRLV =0.12 \pm 0.03$. 
\end{itemize}

While the observed scaling relations of disk galaxies appear to  
be a straightforward result, the lack of correlation between the residuals 
from these relations of $\Vobs$ and $\Rd$ at a given $L$
coupled with their color dependence is more puzzling. In Paper II, 
we argue that the small scatter in the TFR is consistent with being 
a reflection of the natural scatter in halo concentration $c$, as 
measured in cosmological simulations.  This scatter is small because 
it is independent of the scatter in halo spin.  
Other explanations for the small size of the TFR scatter have invoked 
star formation and feedback processes (Eisenstein \& Loeb 1996; Silk 1997; 
Heavens \& Jimenez 1999; FAR00), the small dispersion 
in halo and$/$or disk collapse times (Heavens \& Jimenez 1999; vdB00), and
dynamical response of the halo to the disk assembly (NS00). 
Whether these, and other interpretations for the small scatter in 
the TFR, are all fully independent is still a matter of contention 
that we revisit, along with the interpretation of the major 
observational results above, in Paper II. 

\bigskip
\clearpage

\acknowledgments

We acknowledge useful conversations and suggestions from Eric Bell, 
Hans-Walter Rix, and Marc Verheijen.  Dan Zucker kindly 
helped with the extraction of Petrosian magnitudes from the SDSS/DR1 
database.  Aaron Dutton is also thanked for experimenting with morphological 
classification of 2MASS images.  His work demonstrated the difficulty 
of assigning unambiguous eye-ball morphologies to infrared galaxy images. 
SC wishes to acknowledge his
colleagues on the Shellflow team (Marc Postman, David Schlegel, and 
Michael Strauss) for permission to use previously unpublished results.
SC and LAM acknowledge financial
support from the National Science and Engineering Council of Canada.
AD acknowledges support by the US-Israel Bi-National Science Foundation 
grant 98-00217, the German-Israel Science Foundation grant I-629-62.14/1999,
and NASA ATP grant NAG5-8218.
SC would also like to thank the Max-Planck Institut f{\" u}r Astronomie
in Heidelberg and the Max-Planck Institut f{\" u}r Astrophysik in Munich
for their hospitality while part of this paper was developed. 

This research has made use of (i) the NASA/IPAC Extragalactic
Database (NED) which is operated by the Jet Propulsion Laboratory,
California Institute of Technology, under contract with the National
Aeronautics and Space Administration, as well as NASA's Astrophysics
Data System; (ii), the {\it Two Micron All Sky Survey}, which is a 
joint project of the University of Massachusetts and the Infrared 
Processing and Analysis Center/California Institute of Technology,
funded by the National Aeronautics and Space Administration and
the National Science Foundation; and ($iii$) the {\it Sloan Digital
Sky Survey} (SDSS).  Funding for the creation and distribution of the 
SDSS Archive has been provided by the Alfred P.\ Sloan Foundation, the
Participating Institutions, the National Aeronautics and Space
Administration, the National Science Foundation, the U.S. Department 
of Energy, the Japanese Monbukagakusho, and the Max Planck Society. 
The SDSS Web site is http://www.sdss.org/.
The SDSS is managed by the Astrophysical Research Consortium (ARC)
for the Participating Institutions.  The Participating Institutions 
are the University of Chicago, 
Fermilab, the Institute for Advanced Study, the Japan Participation 
Group, the Johns Hopkins University, Los Alamos National Laboratory,
the Max Planck Institut f\"ur Astronomie (MPIA), the Max Planck 
Institut f\"ur Astrophysik (MPA), New Mexico State University, 
University of Pittsburgh, Princeton University, the United States 
Naval Observatory, and the University of Washington.

\clearpage

\appendix

\section{A General Derivation of the Tully-Fisher relation}{\label{sec:AppendixA}}

Below, we develop the theory for the TFR (see also MMW98; NS00), and
determine the fundamental nature of the TFR.  This is because two
simple predictions for the TFR have been used in the past: one with a log slope 
of $3$ ($-7.5$ in magnitudes), and the other with a log-slope of $4$ ($-10$ 
in magnitudes). Here we compare both predictions and show that they are related 
to one another.  This discussion replicates \se{origin}, but giving more 
detail. 

Let us consider a dark-matter halo of mass $\Mv$ reaching virial 
equilibrium at a time corresponding to the cosmological expansion  
factor $a=(1+z)^{-1}$. 
We assume that the halo properties at that time are relevant for the 
properties of the stellar component that has formed in this halo.

For a body that is dominated by gravitational forces and is approximately 
in steady-state, the virial radius $\Rv$ is defined in  the spirit of the 
spherical-collapse model (Gunn \& Gott 1972) by a given density  contrast 
$\Delta$ relative to the mean universal density at the time of collapse 
(where mass shells are infalling for the first time), namely by $\Mv/\Rv^3 
\prop \Delta\,^{-3}$.  At early  times, when $\omm \simeq 1$,  the relevant 
density contrast is  $\Delta \simeq 178$, while for  the standard $\Lambda$CDM
cosmology (with $\Omega_\Lambda=0.7$ and $\omm =0.3$  today) it rises
to $\Delta \simeq  340$ today (\eg Eke \etal  1998; Peacock 1999). We
ignore the weak redshift dependence of the $\Delta$ factor in the
following expressions\footnote{The maximum correction
is obtained at low redshifts. For example, in the  range $z=0-2$ the
change is  roughly $\Delta \prop  a^{1/2}$, which implies that $a$ in
the following expressions should be replaced by $\Delta^{-1/3} a \prop
a^{5/6}$. This is a weak effect, which becomes even weaker at higher
redshifts  where the bulk of galaxy formation occurs.}.

If disks are embedded in dark halos we can define the total mass of
the system (dark plus baryonic matter) as
\be
 \Mv = {\Rv \Vv^2 \over G}.
\label{eq:totmass}
\ee
Here $G$ is the gravitational constant, and $\Rv$ is defined as
the radius inside of which the average density of the system is 
$\sim 200$ times the critical density of the Universe, which 
at a given $z$ is given by
\be
\rho_{\rm crit} = {3 \, H(z)^2 \over 8 \, \pi \, G}.
\label{eq:critdens}
\ee

One can thus write
\be
 \Rv = {\Vv \over 10 \, H(z)}.
 \label{eq:r_vir}
\ee
Here $\Vv$ is simply the circular velocity at $\Rv$, and $H(z)$
is the Hubble constant at the virialization redshift $z$ of the 
halo.  Note that $a \propto t^{2/3} \propto H(z)^{-2/3}$, for an 
Einstein-de~Sitter cosmology relevant at high redshifts, where $t$ 
is loosely referred to as the time of formation of the halo.
The three {\it virial} quantities at $a$ thus define a one-parameter 
family: $\Mv \prop a^{3/2} \Vv^3 \prop a^{-3} \Rv^3$. 

We now define the disk mass-to-light ratio\footnote{We use the 
subscripts ``d'' and
``h'' to refer to the disk and dark matter halo, respectively.}
\be
 \Upsilond \equiv {\Md \over \Ld},
 \label{eq:upsilond}
\ee
and write
\be
 \Md = \fbar \, \gf \, \Mv.
 \label{eq:galformeff}
\ee
Here $\fbar$ is the baryonic mass fraction of the Universe
($\fbar = \omb/\Omega_0 \lsim 0.1$ for a $\Lambda$CDM
 model), and $\gf$ is the galaxy formation efficiency  
that describes what fraction of the baryonic mass inside the 
halo ultimately ends up in the disk. Depending on the efficiencies of
cooling, star formation and feedback, this can vary anywhere from zero
to unity (see e.g., van den Bosch 2002a)  

Combining equations~(\ref{eq:totmass}) to~(\ref{eq:galformeff}) yields
\be
\label{eq:TFthree}
 \Ld = {\fbar \, \gf \over 10 \, G \, H(z) \, \Upsilond} \, \Vv^3.
\ee
This is a TFR with log-slope $3$ (see also Dalcanton  \etal 1997; 
White 1997; MMW98; van den  Bosch 1998; Syer, Mao, \& Mo 1999).
 
In contrast, several authors in the past have predicted that the TFR
should have a log-slope of $4$ (e.g., Sargent \etal 1977; Aaronson,
Huchra, \& Mould 1979; Salucci, Frenk, \& Persic 1993; Sprayberry
\etal 1995; Zwaan \etal 1995; Stil \& Israel 1998).  Their argument
goes as follows: Assume that the disk surface brightness is described by
an exponential function of the form
\be
\label{eq:surf_disk}
\Sigma(r) = \Sigma_0 \exp(-r/\Rd),
\ee
with $\Rd$ the stellar scale length of the disk. The total luminosity  
of the disk is 
\be
\label{eq:lum_disk}
 \Ld = 2 \pi \Sigma_0 \Rd^2.
\ee
Assuming that one measures the rotation velocity $\Vobs$ at a
radius $r = s \Rd$, and that the gas and stars in the disk move on
circular orbits, 
\be
\label{eq:vel_obs}
 \Vobs^2 = {{G M(r)} \over r},
\ee
with $M(r)$ the total mass within radius $r$. If we define the 
total mass-to-light ratio 
\be
\label{eq:upshat}
\Upstilde(r) \equiv {M(r)\over \Ld} = {\Md(r) + \Mh(r) \over \Ld}
\ee
we can combine equations~(\ref{eq:lum_disk}), (\ref{eq:vel_obs})
and (\ref{eq:upshat}) to obtain

\be
\label{eq:TFfour}
 \Ld = {1 \over 2 \, \pi \, G^2} \; {s^2 \over \Sigma_circ \,
  \Upstilde^2(r)} \Vobs^4
\ee
This is a TFR with a log-slope of $4$, and as long as one
measures the rotation velocity at a constant number of scale lengths,
$s$, the scatter is determined purely by the variation in 
$\Sigma_\circ \Upstilde^2(r)$. This was emphasized by Zwaan \etal (1995),
who, from the finding that both HSB and LSB spirals follow the same TFR,
concluded that LSB galaxies must have much larger values of
$\Upstilde^2(r)$ than their high surface brightness
counterparts.  Note that this ``conspiracy'' does not apply for the TFR
of \Equ{TFthree}, for which the zero-point is independent of the disk's 
surface brightness.

The TFRs in Eqs.~(\ref{eq:TFthree}) and~(\ref{eq:TFfour}) thus
predict a different log-slope.
The cause of this apparent paradox lies in the fact that the surface 
brightness, $\Sigma_0 \propto \Vobs \lambda^{-2}$ (MMW98), and \Equ{TFfour}
reduces to \Equ{TFthree}. Thus, the two theoretical
predictions of the TFR, with log-slopes $3$ and $4$, are directly
related, but only the derivation for \Equ{TFthree} is strictly based 
on dynamics and it is the most fundamental of the two.

\section{SDSS Colors for Galaxy 
     Classification}{\label{sec:AppendixB}}

Accurate optical colors based on digital imaging is available only for the
Shellflow ($VI$) and UMa ($BVI$) samples. The MAT and SCII
galaxy magnitudes were imaged only at $I$-band.  In CR99, we computed
a $B-I$ color term for MAT galaxies using RC3 B-band fluxes but
RC3 magnitudes are inherently uncertain with $\Delta{m}\simeq0.2$ mag.
With the release of SDSS/DR1 data (York \etal 2000; Abazajian \etal 2003), 
we can extract $g$ and $i$ magnitudes 
(as proxy for $V$ and $I$  magnitudes) for galaxies that overlap 
with Shellflow and SCII galaxies.  We do this for Shellflow galaxies
in order to map the SDSS system to the Cousins system, upon which  
Shellflow magnitudes are based.  A final transformation can then be
applied to SCII magnitudes. 
We compare in Figs.~\ref{fig:Shellsloan} \& \ref{fig:SCIIsloan}
instrumental Petrosian  magnitudes from  the SDSS first data release
against instrumental magnitudes for 31 Shellflow and 39 SCII galaxies.
The comparisons avoid any differences due to extinction and k-corrections.  
Besides a zero-point offset, the $g$ and $i$ Petrosian magnitudes 
scale linearly with the $V$ and $I$ Cousins magnitudes in Shellflow 
and SCII.  We compute transformations for the SDSS Petrosian 
magnitudes into Cousins magnitudes for the SCII galaxies that 
overlap with the SDSS/DR1 database. The instrumental ``$V$''
magnitudes for SCII were corrected  the same way as the $I$-band
magnitudes but using an extinction dependence  $A_V=3.24$, instead 
of $A_I=1.96$, and the Poggianti (1997) formulation for the $V$-band
k-correction.

\clearpage
\begin{deluxetable}{lcccccc}
\tablecolumns{7}
\tablenum{1}
\tablewidth{0pc}
\tablecaption{Redshift-Distance Galaxy Surveys\label{tab:tab1}}
\tablehead{
\colhead{Sample} & 
\colhead{$N$} & 
\colhead{Gal. type} & 
\colhead{Phot. bands} & 
\colhead{Mag/Diam. limits} & 
\colhead{Redshift limits} & 
\colhead{Rot. measure} \\
\multicolumn{1}{c}{(1)} & \multicolumn{1}{c}{(2)} & \multicolumn{1}{c}{(3)} & \multicolumn{1}{c}{(4)} & 
\multicolumn{1}{c}{(5)} & \multicolumn{1}{c}{(6)} & \multicolumn{1}{c}{(7)} 
}
\startdata
MAT          & 845 & field   & $I$             &  $D_{\rm ESO} \geq$ 1\farcm7 &  $<7000$ \kms   &  \hi \\
SCII         & 521 & cluster & $I$             &  $12 \leq m_{\rm{I}} \leq 17$&  [5000--19,000] &  \ha, \hi \\
Shell{\sl flow} & 340 & field& $V,I$           &  $m_B \leq 14.5$             &  [4500,7000]    &  \ha \\
UMa          &  44 & cluster & $B,R,I,K$       &  $m_z \leq 14.5$             &  18.6 Mpc       &  \hi \\
\enddata
\end{deluxetable}

\begin{deluxetable}{lrrrr}
\tablecolumns{5}
\tablenum{2}
\tablewidth{0pc}
\tablecaption{Virial Slopes and Correlation Coefficients \label{tab:tab2}}
\tablehead{
\colhead{} & 
\colhead{MAT} & 
\colhead{SCII} & 
\colhead{Shell{\sl flow}} & 
\colhead{UMa}
}
\startdata
$\log\Vobs{(L_I)}$ & $0.31\logL_I~(0.91)$ & $0.31\logL_I~(0.90)$ & $0.31\logL_I~(0.86)$ & $0.32\logL_I~(0.94)$ \\ 
$\log\Rx{(L_I)}$ & $0.31\logL_I~(0.72)$ & $0.34\logL_I~(0.74)$ & $0.35\logL_I~(0.58)$ & $0.18\logL_I~(0.54)$ \\ 
$\log\Rx{(\Vobs)}$ & $0.93\log\Vobs~(0.58)$ & $1.00\log\Vobs~(0.61)$ & $0.64\log\Vobs~(0.42)$ & $0.52\log\Vobs~(0.54)$ \\ \hline

$\log\Vobs{(L_J)}$  & & $0.30\logL_J(0.87)$ & & \\ 
$\log\Re(L_J)$      & & $0.41\logL_J(0.83)$ & & \\ 
$\log\Re(\Vobs)$    & & $1.36\log\Vobs(0.68)$ & & \\ 

\enddata
\end{deluxetable}
\clearpage

\def \partialVRLI {\dlogV{(L_I)} \thinspace / \thinspace\dlogR{(L_I)}}
\def \partialRLVI {\dlogR{(\Vobs)} \thinspace / \thinspace\dlogL{(\Vobs)}}
\def \partialVLRI {\dlogV{(\Rd)} \thinspace / \thinspace\dlogL{(\Rd)}}
\def \partialVRLJ {\dlogV{(L_J)} \thinspace / \thinspace\dlogR{(L_J)}}
\def \partialVLRJ {\dlogV{(\Rd)} \thinspace / \thinspace\dlogL{(\Rd)}}

\begin{deluxetable}{lrrrr}
\rotate
\tablecolumns{5}
\tablenum{3}
\tablewidth{0pc}
\tablecaption{Residual Correlations \label{tab:tab3}}
\tablehead{
\colhead{} & 
\colhead{MAT} & 
\colhead{SCII} & 
\colhead{Shell{\sl flow}} & 
\colhead{UMa}
}
\startdata
$\partialVRLI$ & $-0.04\pm0.02~(-0.26)$           & $-0.07\pm0.04~(-0.16)$            & $-0.08\pm0.04 ~(-0.16)$           & $0.03\pm0.16 ~(+0.11)$ \\
$\partialRLVI$ & $\phantom{-}0.12\pm0.03~(+0.30)$ & $\phantom{-}0.12\pm0.07 ~(+0.33)$ & $\phantom{-}0.15\pm0.06~(+0.35)$  & $0.02\pm0.08 ~(+0.03)$ \\
$\partialVLRI$ & $\phantom{-}0.30\pm0.02~(+0.78)$ & $\phantom{-}0.29\pm0.02 ~(+0.74)$ & $\phantom{-}0.30\pm0.02~(+0.83)$  & $0.29\pm0.06 ~(+0.86)$ \\
               & & & & \\
$\partialVRLJ$ & & $-0.08\pm0.07~(-0.15)$ & & \\
$\partialRLVI$ & & $\phantom{-}0.11\pm0.03 ~(+0.27)$ & & \\
$\partialVLRJ$ & & $\phantom{-}0.26\pm0.03 ~(+0.61)$ & & 

\enddata
\end{deluxetable} 

\clearpage

\begin{figure}
\caption{Scaling relations for MAT galaxies.  Line widths are measured 
 from \ha rotation curves and disk scale lengths are measured from 
 bulge-to-disk decompositions.   Luminosities are computed from 
 fully-corrected $I$-band magnitudes.
 The dashed line is the corresponding fit 
 for the SCII sample.  A significant zero-point difference is seen for
 the respective TFR fits. 
\label{fig:MAT_3rel}}
\end{figure}
\clearpage

\begin{figure}
\caption{Scaling relations for Shellflow galaxies.  Line widths are 
 measured from \ha rotation curves and disk scale lengths are measured 
 from bulge-to-disk decompositions.  Luminosities are 
 computed from fully-corrected $I$-band magnitudes.  
 The dashed line is the corresponding fit for the SCII sample
 (\Fig{SCII_3rel}). 
\label{fig:shell_3rel}}
\end{figure}
\clearpage

\begin{figure}
\caption{Scaling relations for UMa galaxies.  Line widths are measured 
 from \hi synthesis rotation curves and disk scale lengths are from 
 bulge-to-disk decompositions.  Luminosities are 
 computed from fully-corrected $I$-band magnitudes
 The dashed line is the corresponding fit for the SCII 
 sample (\Fig{SCII_3rel}).
\label{fig:UMa_3rel}}
\end{figure}
\clearpage

\begin{figure}
\caption{Combined TF and SL relations for the four data sets
  (SCII, Shellflow, MAT, and UMa) using absolute magnitudes
  (as conventionally plotted by the original authors).  
  Zero-pointing differences clearly exist between each sample. 
\label{fig:allTF}}
\end{figure}
\clearpage

\begin{figure}
\caption{$I$-band magnitude residuals from the TFR for Shellflow galaxies.  
\label{fig:TFres_shell}}
\end{figure}

\begin{figure}
\caption{$I$-band magnitude residuals from the TFR for SCII galaxies. 
\label{fig:TFres_SCII}}
\end{figure}
\clearpage

\begin{figure}
\caption{Magnitude residuals, for the SCII and Shellflow samples,
         as a function of central surface brightness 
     ($\mus \equiv -2.5\log\Sigma_\circ$), disk scale length ($\Rd$), 
     concentration index (CI), red color $V-I$, and optical
         morphological type (see text).  The strongest and tighest 
         correlation with TF residual is against galaxy color. 
\label{fig:TFmag1}}
\end{figure}
\clearpage

\begin{figure}
\caption{Same as \Fig{TFmag1} but for the MAT and UMa samples
         and using a $B-I$ color term.
         CI's values were missing for the MAT galaxies and bulge-to-disk
         ratios, from Byun \& Freeman (1995), were used as substitute.
         The slight trend of TF residual with $\mus$ for the MAT sample
         is due to improper modeling of the galaxy bulge. 
         Galaxy color shows the tightest correlation with TF residuals. 
\label{fig:TFmag2}}
\end{figure}
\clearpage

\begin{figure}
\caption{Correlation of galaxy observables (color, rotation speed, 
         concentration index, and disk scale length) with morphological
         type.  The point types represented different samples
         (green open squares: SCII (with $g-i$ colors); red filled 
          circles: Shellflow (with $V-I$ colors); 
          blue filled triangles: UMa (with $B-I$ colors); 
          the MAT galaxies were omitted for clarity.)  
         Color is found to be most tightly correlated 
         with morphological type.  The degree of concentration, on the
         other hand is a very poor indicator of Hubble type. 
\label{fig:Hubtype}}
\end{figure}
\clearpage

\begin{figure}
\caption{ Scaling relations for a subset of the SCII galaxies 
 ($N=400$) with $J$-band luminosities and effective radii, $\Re$
 in kpc.  
 As in \Fig{SCII_3rel}, the luminosities are not corrected for
 any morphological offset. The point types are a function of $\Re$.  
 We see no striking dependence of TF scatter with $\Re$, other 
 than the smaller systems liying slightly above the mean TFR. 
 The solid lines are our best fits to the data computed as in 
 \se{virial}.  
\label{fig:2MASS_Reff}}
\end{figure}
\clearpage

\begin{figure}
\caption{Same as \Fig{2MASS_Reff}, but with point types as a function 
 of effective surface brightness.  The TFR is clearly independent 
 of surface brightness; 
 the scatter of the SLR is however strongly dominated by it. 
\label{fig:2MASS_SBeff}}
\end{figure}
\clearpage

\begin{figure}
\caption{Same as \Fig{2MASS_Reff}, but with point types as a function 
 of $J-K$ color. The scatter of the TFR is dominated by color.  The 
 SLR shows a weaker, though noticeable, dependence on color. 
\label{fig:2MASS_JmK}}
\end{figure}
\clearpage

\begin{figure}
\caption{Correlation residuals from the mean relation at fixed
 virial quantity for the Shellflow, SCII, MAT, and UMa samples.  
 Symbols are as in \Fig{shell_3rel}. 
 Galaxies of all shapes and types (barred and un-barred, early  
 and later types) obey the same surface brightness independent
 scaling relations. 
 The residual distribution for $\partialvr$
 of all three samples is flat for barred and un-barred 
 galaxies of all Hubble types and luminosities.  The middle figures 
 show no correlation between $L$ and $R$, and the figures on the right 
 side recast the TFR in its differential form. 
\label{fig:3res}}
\end{figure}
\clearpage

\begin{figure}
\caption{Correlation residuals for the SCII sample with 2MASS $J$-band 
 luminosities and effective radii, separated by $J-K$ color to show the 
 differences from red and bluer.  The inscriptions in the bottom 
 windows show the mean Pearson coefficients and residual slopes for
 all the colors combined. 
 Redder/bluer galaxies lie slightly above/below the null line in the TF/SL 
 relations (left panels).
\label{fig:2MASS_3res}}
\end{figure}
\clearpage

\begin{figure}
\caption{TF/SL correlation residuals at fixed virial quantity for the 
 Shellflow, SCII, and UMa samples.  Symbols are as in \Fig{SCII_3rel}.
 The solid and long dashed line in the upper panel 
 show the best fit and the statistically acceptable range of 
 these slopes (at the 95\% confidence level), respectively. 
 The best fit is given by $\partialVRL = -0.07 \pm 0.03$. 
 The short dashed line in the upper panel shows the 
 maximal disk solution [$\partialVRL=-0.5$]. 
 The residual dependences on color (bottom) use the 
 Shellflow $V-I$ data, the SLOAN $g$ and $i$ magnitudes transformed 
 according to Figs.~\ref{fig:Shellsloan} \& \ref{fig:SCIIsloan} for 
 a subset of 39 SCII galaxies (see Appendix A), and the $B-I$ colors 
 for the MAT and UMa samples scaled to match the Shellflow $V-I$ range. 
\label{fig:dvdr}}
\end{figure}
\clearpage

\begin{figure}
\caption{Correlation residuals for the SCII sample with $J$-band 
 photometry.  The line types are as in \Fig{dvdr}.  The best fit 
 is given by $\partialVRL = -0.08 \pm 0.07$.  The residual 
 dependences on $J-K$ color are shown in the bottom panels 
 with the same line types as above for the right lower panels. 
 Those fits are given by $\partial\log{V(L)}/\partial(J-K)=0.09 \pm 0.04$
 and $\partial\log{R(L)}/\partial(J-K)=-0.19 \pm 0.15$.
 The point types in the top figure refer to different colors 
 (same notation as in \Fig{2MASS_JmK}) and the point types 
 in the four bottom panels refer to different surface brightness 
 levels (as in \Fig{2MASS_SBeff}).
\label{fig:2MASSdvdr}}
\end{figure}
\clearpage

\begin{figure}
\caption{Comparison of raw SLOAN (Petrosian) vs Shellflow (Cousins) 
 magnitudes. This confirms that the two sets are derivable from 
 one another, independent of a color term at least over the small 
 Shellflow magnitude range.
\label{fig:Shellsloan}}
\end{figure}
\clearpage

\begin{figure}
\caption{Comparison of raw SLOAN (Petrosian) vs SCII (Cousins) 
 magnitudes.  The linear transformation of magnitudes holds over 
 nearly 3 magnitudes, independent of a color term.
\label{fig:SCIIsloan}}
\end{figure}
\clearpage

\end{document}